\documentclass[preprintnumbers,showpacs,11pt,amssymb,nofootinbib]{revtex4}

\usepackage[T1]{fontenc} 

\usepackage{amssymb}

\usepackage{bm}

\newcommand{\be}{\begin{equation}}
\newcommand{\ee}{\end{equation}}
\newcommand{\beqq}{\setlength\arraycolsep{2pt}\begin{eqnarray}}
\newcommand{\eeqq}{\vspace{0cm} \end{eqnarray}}
\newcommand{\bea}{\begin{eqnarray}}
\newcommand{\eea}{\end{eqnarray}}

\newcommand{\lambdab}{\stackrel{\neg}{\lambda}}
\newcommand{\Phib}{\stackrel{\neg}{\Phi}}
\newcommand{\e}{\textrm{e}}

\begin{document}

\title{Exact solutions to Elko spinors in spatially flat Friedmann-Robertson-Walker spacetimes}

\author{J. M. Hoff da Silva} \email{hoff@feg.unesp.br}
\author{S. H. Pereira} \email{shpereira@gmail.com}

\affiliation{Faculdade de Engenharia de Guaratinguet\'a -- UNESP - Univ. Estadual Paulista \\ Departamento de F\'isica e Qu\'imica\\ Av. Dr. Ariberto Pereira da Cunha 333 - Pedregulho\\
12516-410 -- Guaratinguet\'a, SP, Brazil}


\begin{abstract}In this paper we present exact solutions to the so-called Elko spinors for three models of expanding universe, namely the de Sitter, linear and the radiation type evolution. The study was restricted to flat, homogeneous and isotropic Friedmann-Robertson-Walker backgrounds. Starting with an Elko spinor we present the solutions for these cases and compare to the case of Dirac spinors. Besides, an attempt to use Elko spinors as a dark energy candidate in the cosmological context is investigated.
\end{abstract}

\maketitle

\section{Introduction}

Exact solutions to the Dirac equation in curved spacetime is of considerable interest in cosmology and astrophysics, where gravity is believed to play a dominant role in determining the behavior of spin-1/2 particles. A general discussion on the interaction of massless neutrinos and spherically symmetric gravitational fields was performed by Brill and Wheeler \cite{brill} in 1957. In the 1970s the phenomenon of particle production in curved spacetime was investigated by Parker \cite{parker} and in 1974 Hawking discovered the effect of black hole evaporation \cite{hawking1,halking2}, an appropriate example regarding the importance of strong gravitational fields in quantum mechanical processes. Also, the study of the hydrogen atom energy spectrum in curved spacetime was presented by Audretsch and Sch\"afer \cite{schafer} and was also studied by Parker in 1980 \cite{atomParker}.

Finding exact and analytic solutions of the Dirac equation in curved backgrounds is always a hard task. Some exact solutions have been reported in the middles of 1980 \cite{after1,after2}. In 1987 Barut and Duru \cite{barut} provided an exact solution for the Dirac equation for a spatially flat Friedmann-Robertson-Walker spacetime in three meaningful models of expanding universes, based on the spin connection point of view. Exact solutions of the Dirac equation in open and closed Friedmann-Robertson-Walker spaces were presented in subsequent years for both massive and massless case \cite{nonflat1,nonflat2,nonflat3,nonflat4,nonflat5}. Solutions for Kasner spacetime was obtained by Srivastava \cite{sushil} and for an anisotropic Bianchi type VI background was presented by Portugal \cite{portugal}. 

In this work we aim to investigate exact solutions for Elko spinors whose dynamics is taken in curved spacetime. More precisely, we study the solutions for the aforementioned spinor field in spatially flat Friedmann-Robertson-Walker spacetimes. Elko spinor fields were introduced  in \cite{AHL1,AHL2} as a possible generalization of Majorana spinor fields. The main property defining Elko spinors is that they are eigenspinors of the charge conjugator operator, making them neutral under electromagnetic interactions by construction. Since the introduction of the Elko spinors, modifications and improvements have been accomplished. The final form for the spinor and its corresponding quantum theory may be found in \cite{AHL3}. There are several works considering Elko spinor fields in the context of curved spacetimes and cosmology. The study of Elko spinors with a possible coupling with torsion fields is presented in \cite{BOE1,BOE2}, as well as its impact on Cosmic Microwave Background anisotropies \cite{BOE5} and its relation to the cosmological principle \cite{FABBRI}. Following this reasoning, important consequences of dark spinor models to inflation are studied in \cite{BOE3,BOE4} and interesting solutions where the dark spinor field leads to slow roll and fast roll de Sitter solutions are presented in \cite{BOE6}. Scalar and tensor cosmological perturbations are discussed in \cite{GREDAT,BASAK} and dark spinor models as a candidate of dark energy are investigated in \cite{BOE7,WEI}, as well as the cosmological coincidence problem.

As remarked, the endeavour on finding exact solutions of spinor fields in curved spacetimes is important in several contexts. The possibility, raised in the Elko formalism, of understanding this spinor field as a candidate to dark matter (for instance, along with the fact that Elko has a peculiar dynamics) certainly highlights the relevance of studying exact solutions for the Elko dynamics in physically important spacetimes. Furthermore, it is also a robust starting point to investigate Elko particle production in curved backgrounds \cite{TLV}.     

This paper is organized as follows: in the next Section we give a tutorial and short review about the main aspects of Elko spinor fields. In Section III we study exact solutions of Elko dynamics in three different cosmological expanding spacetimes, namely: the de Sitter one, a linear expansion and the radiation dominated universe. Section IV is reserved to the investigation of the obtained solutions in cosmology. In the final Section we conclude, comparing the obtained solutions with the usual case of Dirac spinors.  

\section{Elko spinor fields}

In this Section we shall review some important aspects concerning Elko spinor fields and its dynamics \cite{AHL1, AHL3}. As mentioned in the Introduction, the very relation defining Elko spinor fields is given by
\be
C\lambda =\pm \lambda, \label{E1}
\ee where $C$ stands for the charge conjugator operator. Hence, $\lambda$ is an eigenspinor of $C$. By solving Eq. (\ref{E1}), it is possible to recast the spinors as self-conjugate ($+$ sign in (\ref{E1})) $\lambda_{\{+,-\}}^S$, $\lambda_{\{-,+\}}^S$, and anti-self-conjugate ($-$ sign in (\ref{E1})) $\lambda_{\{+,-\}}^A$, $\lambda_{\{-,+\}}^A$. They are given explicitly by 
\begin{eqnarray}
\lambda^{S}_{\{+,-\}}(\vec{0})=\left(
                              \begin{array}{cc}
                                +\sigma_{2}[\phi_{L}^{-}(\vec{0})]^{\ast} \\
                                \phi_{L}^{-}(\vec{0}) \\
                              \end{array}
                            \right),\nonumber \\
                            \lambda^{S}_{\{-,+\}}(\vec{0})=\left(
                                                    \begin{array}{cc}
                                                      +\sigma_{2}[\phi_{L}^{+}(\vec{0})]^{\ast} \\
                                                      \phi_{L}^{+}(\vec{0}) \\
                                                    \end{array}
                                                  \right),\nonumber\\
                                                  \lambda^{A}_{\{+,-\}}(\vec{0})=\left(
                              \begin{array}{cc}
                                -\sigma_{2}[\phi_{L}^{-}(\vec{0})]^{\ast} \\
                                \phi_{L}^{-}(\vec{0}) \\
                              \end{array}
                            \right),\nonumber \\
                            \lambda^{A}_{\{-,+\}}(\vec{0})=-\left(
                              \begin{array}{cc}
                                -\sigma_{2}[\phi_{L}^{+}(\vec{0})]^{\ast} \\
                                \phi_{L}^{+}(\vec{0}) \\
                              \end{array}
                            \right),\label{e4}
\end{eqnarray} with phases adopted such that
\begin{eqnarray}
\phi^{+}_{L}(\vec{0})=\sqrt{m}\left(
                           \begin{array}{c}
                             \cos(\theta/2)e^{-i\phi/2} \\
                             \sin(\theta/2)e^{i\phi/2} \\
                           \end{array}
                         \right)\label{e5}
\end{eqnarray} and
\begin{eqnarray}
\phi^{-}_{L}(\vec{0})=\sqrt{m}\left(
                           \begin{array}{c}
                             -\sin(\theta/2)e^{-i\phi/2} \\
                              \cos(\theta/2)e^{i\phi/2} \\
                           \end{array}
                         \right).\label{e6}
\end{eqnarray}

The equations above are valid in the rest frame $(\vec{k}=\vec{0})$, therefore the expressions for arbitrary momenta are obtained by a simple boost. The parameter $m$ denotes the spinor field mass, $\sigma_2$ is the usual Pauli matrix, and the momentum parametrization is given by $\hat{k}=(\sin\theta \cos\phi,\sin\theta \sin\phi, \cos\theta)$. It is remarkable that $-i\sigma_2[\phi_{L}^{\pm}(\vec{0})]^*$ and $\phi_L^{\pm}(\vec{0})$ have opposite helicities. It means that Elko spinor fields belong to the $\Big(\frac{1}{2},0\Big)\oplus \Big(0,\frac{1}{2}\Big)$ representation space.

The dual spinor associated to $\lambda^{S/A}$ can be obtained in a very judicious way, by demanding that the product $\lambdab \lambda$, being $\lambdab$ the dual, remains invariant under Lorentz transformations \cite{PLBAHL}. The result reads
\be
\lambdab_{\{\mp,\pm\}}^{S/A}(\vec{k})=\pm i\Big[\lambda_{\{\pm,\mp\}}^{S/A}(\vec{0})\Big]^{\dagger}\gamma^{0}.\label{e7}
\ee With the aid of Eq. (\ref{e7}) it is possible to write down the spin sums  
\begin{eqnarray}
\sum_{\kappa}\lambda_{\kappa}^{S}\lambdab_{\kappa}^{S}=+m[\mathbb{I}+\mathcal{G}(\phi)],\nonumber\\
\sum_{\kappa}\lambda_{\kappa}^{A}\lambdab_{\kappa}^{A}=-m[\mathbb{I}-\mathcal{G}(\phi)],\label{e8}
\end{eqnarray} where $\mathcal{G}(\phi)$ is given by \cite{PLBAHL}
\begin{eqnarray}
\mathcal{G}(\phi)=\gamma^{5}(\gamma_{1}\sin\phi-\gamma_{2}\cos\phi).\label{e9}
\end{eqnarray} 

In order to unveil the Elko quantum dynamics we need an approach different from the usual textbook ones, inasmuch as we do not know a priori what Lagrangian must be associated to the Elko spinor. The first hint towards its dynamics comes from the following algebraic relation
\be
(\gamma_{\mu}k^{\mu}\delta_{\alpha}^{\beta}\pm im \mathbb{I}\varepsilon_{\alpha}^{\beta})\lambda_{\beta}^{S/A}({\vec{k}})=0,\label{e14}
\ee which can be obtained by applying $\gamma_\mu k^\mu$ to $\lambda_{\beta}^{S/A}(\vec{k})$. From Eq. (\ref{e14}) it is straightforward to see that the application of $\gamma^\nu k_\nu$ from the left leads to 
\be
(\gamma^\nu\gamma^\mu k_\mu k_\nu-m^2)\lambda^{S/A}_{\{\mp,\pm\}}=0,\label{KGE}
\ee which, by means of $\{\gamma^\mu,\gamma^\nu\}=2\eta^{\mu\nu}$, leads to the Klein-Gordon equation in the momentum. In the following we shall derive the Klein-Gordon equation of the Elko spinor field by a more precise argument.

(\ref{e14}) is a Dirac-like equation. It is obviously different from the Dirac equation, but they share the covariant structure. Hence, denoting a spinorial transformation as $\lambda'=S \lambda$ (assuming that $\lambda$ belongs to a linear representation of, at least, a subgroup of the Lorentz group), and studying the transformation of the expression (\ref{e14}) we arrive at the same covariance condition of the standard Dirac equation $S\gamma^\nu S^{-1}\Lambda^{\mu}_{\;\;\nu}=\gamma^\mu$. Therefore, as in the Dirac case, the field $\lambda$ is not unitarily transformed and cannot be associated to a quantum state. Thus, quantization is necessary. 

By keeping some recurrence with the usual spinorial case, we may associate a quantum field by 
\begin{eqnarray}
\eta(x)=\int \frac{d^{3}k}{(2\pi)^{3}}\frac{1}{\sqrt{2mE(\vec{k})}}\sum_{\alpha}[c_{\alpha}(\vec{k})\lambda^{S}_{\alpha}(\vec{k})
e^{-ik_{\mu}x^{\mu}}+c_{\alpha}^{\dagger}(\vec{k})\lambda^{A}_{\alpha}(\vec{k})e^{+ik_{\mu}x^{\mu}}],\label{e12}
\end{eqnarray} where $c_{\alpha}^{\dagger}(\vec{k})$ and $c_{\alpha}(\vec{k})$ are the creation and annihilation operators, respectively, satisfying the usual fermionic anti-commutation relations. The quantum dual may be obtained in a rather similar fashion. With the quantum fields at hands we may evaluate the Feynman-Dyson propagator, given by 
\be
S_{FD}(x'-x)=i\langle |\mathcal{T}\big(\eta(x')\stackrel{\neg}{\eta}(x)\big)|\rangle,
\ee where $\mathcal{T}$ is the time ordering operator. The calculation is a little tricky due to the subtle aspects of the field. The final result reads \cite{AHL1,AHL2,AHL3}
\be
S_{FD}(x'-x)=-\int \frac{d^4k}{(2\pi)^4}e^{-ik^\mu(x'_\mu-x_\mu)}\Bigg[\frac{1}{k_\mu k^\mu-m^2}\Bigg],
\ee hence the Elko spinor field must respect (only) the Klein-Gordon Lagrangian, i.e., it has mass dimension one. If we keep ourselves on power counting arguments, then the only perturbatively renormalizable possible terms are the mass one, the self (quartic) interaction $(\lambdab \lambda)^2$ and the coupling with a scalar field. 

In the following we shall investigate the exact solutions for the Elko spinor field in the context of physically relevant expanding spacetimes.

\section{The Elko spinor equation in expanding spacetimes}

By the reasons exposed in the previous Section, the Elko spinor action in the curved spacetime is given by:
\be
S={1\over 2}\int\sqrt{-g}\bigg({1\over 2}g^{\mu\nu}\big(\nabla_\mu \lambdab \nabla_\nu\lambda + \nabla_\nu \lambdab \nabla_\mu\lambda \big) - V(\lambdab \lambda)\bigg)d^4x\,,\label{action}
\ee
where $V(\lambdab \lambda)$ is the potential and $g\equiv$ det$g_{\mu\nu}$. The covariant derivatives acting on the Elko spinors are $\nabla_\mu \lambdab = \partial_\mu \lambdab + \lambdab \Gamma_\mu$ and $\nabla_\mu \lambda = \partial_\mu \lambda -\Gamma_\mu \lambda$, where $\Gamma_\mu$ are the spin connections. 

The metric in a spatially flat, homogeneous and isotropic Friedmann-Robertson-Walker expanding universe is given by
\be
ds^2=dt^2 - a^2(t)(dx^2+dy^2+dz^2)\,,\label{ds}
\ee
thus
\be
g_{\mu\nu}=diag(1,-a^2,-a^2,-a^2)\,,\;\;\;g^{\mu\nu}=diag(1,-1/a^2,-1/a^2,-1/a^2)\,,\label{metric}
\ee
with $g^{\mu\alpha}g_{\alpha\nu}=\delta^\mu_\nu$ and $\sqrt{-g}=a^3$. In order to satisfy the defining equations $\gamma^\mu\gamma^\nu + \gamma^\nu\gamma^\mu = 2g^{\mu\nu}$ with respect to the metric (\ref{metric}), the Dirac matrices $\gamma^\mu(x)$ are
\be
\gamma^0(t)=\gamma_0\,,\hspace{1cm}\gamma^i(t)=-{1\over a(t)}\gamma_i\,,\;\;\;i=1,\,2,\,3\,,
\ee
where $\gamma_\mu$ (lower index) denotes the standard Dirac matrices in the Minkowiski space. The spin connections $\Gamma_\mu$ can be determined as $\Gamma_0=0$ and $\Gamma_i={\dot{a}\over 2} \gamma_0\gamma_i$, where a dot denotes a time derivative.

Taking the potential of the form $V={1\over 2}m^2\lambdab \lambda$, the Elko Lagrangian density can be written as
\be
\mathcal{L}={1\over 2}\sqrt{-g}\bigg[g^{\mu\nu}(\nabla_\mu \lambdab \nabla_\nu \lambda) -m^2\lambdab \lambda\bigg]\,.\label{lagrangian}
\ee
The equations of motion follow from a principle of least action. For the spinor $\lambda$, for instance, we have
\be
\partial_\alpha\bigg[\sqrt{-g}g^{\alpha\nu}(\partial_\nu \lambda - \Gamma_\nu \lambda)\bigg]+\sqrt{-g}\bigg[g^{\mu\nu}(\Gamma_\mu\Gamma_\nu \lambda - \Gamma_\mu \partial_\nu \lambda) + m^2 \lambda\bigg]=0\,,
\ee
and the corresponding equation of motion taking into account the metric (\ref{metric}) is
\be
\ddot{\lambda} + 3 \bigg({\dot{a}\over a}\bigg)\dot{\lambda}-{1\over a^2}\partial_i^2\lambda - {3\over 4}\bigg({\dot{a}\over a}\bigg)^2\lambda + m^2 \lambda + {\dot{a}\over a^2}\gamma_0 \gamma_i (\partial_i \lambda)=0\,,\label{eqmotion}
\ee
where we have used $\Gamma_i\Gamma_i ={\dot{a}^2\over 4}$I. The corresponding equation for $\lambdab$ is
\be
\ddot{\lambdab} + 3 \bigg({\dot{a}\over a}\bigg)\dot{\lambdab}-{1\over a^2}\partial_i^2\lambdab - {3\over 4}\bigg({\dot{a}\over a}\bigg)^2\lambdab + m^2 \lambdab - {\dot{a}\over a^2}(\partial_i \lambdab)\gamma_0 \gamma_i =0\,.\label{eqmotionb}
\ee
These equations are the generalization of the corresponding equations of motion obtained in \cite{BOE3,BOE6} for the scalar part of the Elko field, including the non-homogeneous terms of the type $\partial_i\lambda$.

Since $a$ is a function of $t$ only, we can set
\begin{eqnarray}
\lambda(\vec{x},t)=N{\e^{i\vec{k}\cdot\vec{x}}\over a(t)^{3/2}}
\left(\begin{array}{c}
\Phi_I(\vec{k},t)\\ \Phi_{II}(\vec{k},t)
\end{array}\right),\label{geral}
\end{eqnarray}
where $N$ is a normalization constant. It is a fairly simple exercise to constraint the $\Phi_{I,II}$ components of (\ref{geral}) by means of the eigenspinor equation (\ref{E1}) in the rest frame, in order to ensure the spinor in question as an Elko spinor field indeed. To fix ideas, let us call $\Phi_I^T=(\phi_1(t)\alpha, \phi_2(t)\beta)$, $\Phi_{II}^T=(\phi_3(t)\gamma, \phi_4(t)\delta)$, being $\alpha, \beta, \gamma, \delta$ constants, and work with the positive sign of (\ref{E1}). The result is
\begin{eqnarray}
\Phi_I(\vec{k},t)=
\left(\begin{array}{c}
\phi_1(t)\alpha(\vec{k})\\ i\phi_3^*(t)\gamma^*(\vec{k})
\end{array}\right),\hspace{1cm}
\Phi_{II}(\vec{k},t)=
\left(\begin{array}{c}
\phi_3(t)\gamma(\vec{k})\\ -i\phi_1^*(t)\alpha^*(\vec{k})
\end{array}\right).
\end{eqnarray}\label{lambda} 
The functions $\Phi_I$ and $\Phi_{II}$ will satisfy the following equation, 
\be
\ddot{\Phi}_{I,II} +\Bigg[{\vec{k}^2\over a^2}+m^2 - {3\over 2}\bigg({\dot{a}\over a}\bigg)^2-{3\over 2}{\ddot{a}\over a}\Bigg]\Phi_{I,II} \pm i{\dot{a}\over a^2}\vec{k}\cdot \vec{\sigma} \,\Phi_{I,II}=0\,,\label{eqmotion1}
\ee
where the plus and minus signal stands for $\Phi_I$ and $\Phi_{II}$, respectively. It is interesting to note from this equation that the corresponding equations for $\Phi_I$ and $\Phi_{II}$ are decoupled, but due to the last term the equations for $\phi_1(t)$ and $\phi_3^*(t)$ are coupled, and the same happens for $\phi_3(t)$ and $\phi_1^*(t)$. 

To solve this system of coupled equations we make the decomposition $\phi_1(t)=\phi_{1R}(t)+i\phi_{1I}(t)$ and $\phi_3(t)=\phi_{3R}(t)+i\phi_{3I}(t)$, where $\phi_{1R}, \,\phi_{3R}$ stands for the real part of $\phi_1$ and $\phi_3$, respectively, and $\phi_{1I}, \,\phi_{3I}$ for the imaginary part. Substituting in (\ref{eqmotion1}) we have the four coupled differential equations:

\be
\ddot{\phi}_{1}(t) +\Bigg[{\vec{k}^2\over a^2}+m^2 - {3\over 2}\bigg({\dot{a}\over a}\bigg)^2-{3\over 2}{\ddot{a}\over a}\Bigg]\phi_{1}(t) + i{\dot{a}\over a^2} \bigg(k_3\phi_1(t) + {i\gamma^* \over \alpha}k_- \phi_3^*(t)\bigg)=0\,,\label{eqmotion21}
\ee
\be
\ddot{\phi}_{3}^*(t) +\Bigg[{\vec{k}^2\over a^2}+m^2 - {3\over 2}\bigg({\dot{a}\over a}\bigg)^2-{3\over 2}{\ddot{a}\over a}\Bigg]\phi_{3}^*(t) + i{\dot{a}\over a^2} \bigg({\alpha\over i\gamma^*}k_+\phi_1(t) - k_3 \phi_3^*(t)\bigg)=0\,,\label{eqmotion22}
\ee
\be
\ddot{\phi}_{3}(t) +\Bigg[{\vec{k}^2\over a^2}+m^2 - {3\over 2}\bigg({\dot{a}\over a}\bigg)^2-{3\over 2}{\ddot{a}\over a}\Bigg]\phi_{3}(t) - i{\dot{a}\over a^2} \bigg(k_3\phi_3(t) - {i\alpha^* \over \gamma}k_- \phi_1^*(t)\bigg)=0\,,\label{eqmotion23}
\ee
\be
\ddot{\phi}_{1}^*(t) +\Bigg[{\vec{k}^2\over a^2}+m^2 - {3\over 2}\bigg({\dot{a}\over a}\bigg)^2-{3\over 2}{\ddot{a}\over a}\Bigg]\phi_{1}^*(t) - i{\dot{a}\over a^2} \bigg({\gamma\over -i\alpha^*}k_+\phi_3(t) - k_3 \phi_1^*(t)\bigg)=0\,,\label{eqmotion24}
\ee
with $k_\pm = k_1 \pm i k_2$. 

\vspace{0.5cm}

For the anti-self-conjugate spinor $\lambdab$ we use the definition (\ref{e7}). The $\pm i$ factor is irrelevant to equation (\ref{eqmotionb}), thus we set: 
\begin{eqnarray}
\lambdab(\vec{x},t)=N{\e^{-i\vec{k}\cdot\vec{x}}\over a(t)^{3/2}} \bigg\{ \Phib_I(\vec{k},t)\,,\;\;\; \Phib_{II}(\vec{k},t) \bigg\} \label{geralb}
\end{eqnarray}
where 
\begin{eqnarray}
\Phib_I(\vec{k},t)= \bigg\{ \phi_3^*(t)\gamma^*(\vec{k})\,,\;\;\; i\phi_1(t)\alpha(\vec{k})\bigg\}\,,
\end{eqnarray}
\begin{eqnarray}
\Phib_{II}(\vec{k},t)= \bigg\{ \phi_1^*(t)\alpha^*(\vec{k})\,,\;\;\; -i\phi_3(t)\gamma(\vec{k}) \bigg\}\,.
\end{eqnarray} The full set of coupled equations for this case is

\be
\ddot{\phi}_{3}^*(t) +\Bigg[{\vec{k}^2\over a^2}+m^2 - {3\over 2}\bigg({\dot{a}\over a}\bigg)^2-{3\over 2}{\ddot{a}\over a}\Bigg]\phi_{3}^*(t) - i{\dot{a}\over a^2} \bigg({\alpha\over i\gamma^*}k_+\phi_1(t) - k_3 \phi_3^*(t)\bigg)=0\,,\label{eqmotion22b}
\ee
\be
\ddot{\phi}_{1}(t) +\Bigg[{\vec{k}^2\over a^2}+m^2 - {3\over 2}\bigg({\dot{a}\over a}\bigg)^2-{3\over 2}{\ddot{a}\over a}\Bigg]\phi_{1}(t) - i{\dot{a}\over a^2} \bigg(k_3\phi_1(t) + {i\gamma^* \over \alpha}k_- \phi_3^*(t)\bigg)=0\,,\label{eqmotion21b}
\ee
\be
\ddot{\phi}_{1}^*(t) +\Bigg[{\vec{k}^2\over a^2}+m^2 - {3\over 2}\bigg({\dot{a}\over a}\bigg)^2-{3\over 2}{\ddot{a}\over a}\Bigg]\phi_{1}^*(t) + i{\dot{a}\over a^2} \bigg({\gamma\over -i\alpha^*}k_+\phi_3(t) - k_3 \phi_1^*(t)\bigg)=0\,,\label{eqmotion24b}
\ee
\be
\ddot{\phi}_{3}(t) +\Bigg[{\vec{k}^2\over a^2}+m^2 - {3\over 2}\bigg({\dot{a}\over a}\bigg)^2-{3\over 2}{\ddot{a}\over a}\Bigg]\phi_{3}(t) + i{\dot{a}\over a^2} \bigg(k_3\phi_3(t) - {i\alpha^* \over \gamma}k_- \phi_1^*(t)\bigg)=0\,.\label{eqmotion23b}
\ee


\subsection{Case $a(t)=a_0\,e^{Ht}$}

For the case $a(t)=a_0\,\e^{Ht}$, which represents an inflationary universe or a de Sitter evolution, the coupled equations (\ref{eqmotion21}) - (\ref{eqmotion24}) have the following solutions in terms of the Whittaker $M_{\mu,\nu}(z)$ and $W_{\mu,\nu}(z)$ functions \cite{math}:
\beqq
\phi_1(t) = {2\e^{{1\over 2}Ht}\over \alpha \gamma k_+ + \alpha^*\gamma^* k_-}&\bigg[&\Big(c_5\alpha^*\gamma^*k_- +i c_1 |\gamma|^2(k_3+k)\Big)M_{+1/2,\nu}(z)\nonumber\\&+& \Big(c_7\alpha^*\gamma^*k_- +ic_3 |\gamma|^2(k_3+k)\Big)W_{+1/2,\nu}(z)\nonumber\\ &+&\Big(c_6\alpha^*\gamma^*k_- +ic_2 |\gamma|^2(k_3-k)\Big)M_{-1/2,\nu}(z)\nonumber\\&+& \Big(c_8\alpha^*\gamma^*k_- + ic_4|\gamma|^2(k_3-k)\Big)W_{-1/2,\nu}(z)\,\,\bigg]\,,\label{eq316}
\eeqq
\beqq
\phi_3(t) = {2\e^{{1\over 2}Ht}\over \alpha \gamma k_+ + \alpha^*\gamma^* k_-}&\bigg[&\Big(c_1\alpha^*\gamma^*k_- - ic_5 |\alpha|^2(k_3-k)\Big)M_{+1/2,\nu}(z)\nonumber\\&+& \Big(c_3\alpha^*\gamma^*k_- - ic_7|\alpha|^2(k_3-k)\Big)W_{+1/2,\nu}(z)\nonumber\\ &+&\Big(c_2\alpha^*\gamma^*k_- - ic_6|\alpha|^2(k_3+k)\Big)M_{-1/2,\nu}(z)\nonumber\\&+& \Big(c_4\alpha^*\gamma^*k_- -ic_8|\alpha|^2(k_3+k)\Big)W_{-1/2,\nu}(z)\,\,\bigg]\,,\label{eq317}
\eeqq
where $c_i$ ($i=1, \,2,\,\dots ,8$) are integration constants, $k\equiv |\vec{k}|$, $\nu=\sqrt{3-m^2/H^2}$ and $z=2ik/(Ha_0\e^{Ht})$. 

Finally, we can write the four independent solutions as:
\begin{eqnarray}
\lambda_1(\vec{x},t)={2\e^{i\vec{k}\cdot\vec{x}}\over a_0^{3/2} \e^{Ht}}{M_{+1/2,\nu}(z) \over [\alpha \gamma k_+ + \alpha^*\gamma^* k_-]}
   \left(\begin{array}{c}
   c_5|\alpha|^2\gamma^*k_- + i c_1  |\gamma|^2\alpha(k_3+k)\\
   -c_5|\alpha|^2\gamma^*(k_3-k) + ic_1 |\gamma|^2\alpha k_+\\
   c_1|\gamma|^2\alpha^*k_- -i c_5 |\alpha|^2\gamma(k_3-k)\\
   -c_1|\gamma|^2\alpha^* (k_3+k) - ic_5 |\alpha|^2\gamma k_+
\end{array}\right)\,,\label{sol1}
\end{eqnarray}
\begin{eqnarray}
\lambda_2(\vec{x},t)={2\e^{i\vec{k}\cdot\vec{x}}\over a_0^{3/2} \e^{Ht}}{W_{+1/2,\nu}(z) \over [\alpha \gamma k_+ + \alpha^*\gamma^* k_-]}
   \left(\begin{array}{c}
   c_7|\alpha|^2\gamma^*k_- + i c_3  |\gamma|^2\alpha(k_3+k)\\
   -c_7|\alpha|^2\gamma^*(k_3-k) + ic_3 |\gamma|^2\alpha k_+\\
   c_3|\gamma|^2\alpha^*k_- -i c_7 |\alpha|^2\gamma(k_3-k)\\
   -c_3|\gamma|^2\alpha^* (k_3+k) - ic_7 |\alpha|^2\gamma k_+
\end{array}\right)\,, \label{sol2}
\end{eqnarray}
\begin{eqnarray}
\lambda_3(\vec{x},t)={2\e^{i\vec{k}\cdot\vec{x}}\over a_0^{3/2} \e^{Ht}}{M_{-1/2,\nu}(z) \over [\alpha \gamma k_+ + \alpha^*\gamma^* k_-]}
   \left(\begin{array}{c}
   c_6|\alpha|^2\gamma^*k_- + i c_2  |\gamma|^2\alpha(k_3-k)\\
   -c_6|\alpha|^2\gamma^*(k_3+k) + ic_2 |\gamma|^2\alpha k_+\\
   c_2|\gamma|^2\alpha^*k_- -i c_6 |\alpha|^2\gamma(k_3+k)\\
   -c_2|\gamma|^2\alpha^* (k_3-k) - ic_6 |\alpha|^2\gamma k_+
\end{array}\right)\,, \label{sol3}
\end{eqnarray}
\begin{eqnarray}
\lambda_4(\vec{x},t)={2\e^{i\vec{k}\cdot\vec{x}}\over a_0^{3/2} \e^{Ht}}{W_{-1/2,\nu}(z) \over [\alpha \gamma k_+ + \alpha^*\gamma^* k_-]}
   \left(\begin{array}{c}
   c_8|\alpha|^2\gamma^*k_- + i c_4  |\gamma|^2\alpha(k_3-k)\\
   -c_8|\alpha|^2\gamma^*(k_3+k) + ic_4 |\gamma|^2\alpha k_+\\
   c_4|\gamma|^2\alpha^*k_- -i c_8 |\alpha|^2\gamma(k_3+k)\\
   -c_4|\gamma|^2\alpha^* (k_3-k) - ic_8 |\alpha|^2\gamma k_+
\end{array}\right)\,. \label{sol4}
\end{eqnarray}

For the anti-self-conjugate spinor $\lambdab$ we have:
\begin{eqnarray}
\lambdab_1(\vec{x},t)={2\e^{-i\vec{k}\cdot\vec{x}}\over a_0^{3/2} \e^{Ht}}{M_{+1/2,\nu}(z) \over [\alpha \gamma k_+ + \alpha^*\gamma^* k_-]}
   &\Big\{&
   c_5|\alpha|^2\gamma^*k_- + i c_1  |\gamma|^2\alpha(k_3-k)\,,\nonumber\\
   &-&c_5|\alpha|^2\gamma^*(k_3+k) + ic_1 |\gamma|^2\alpha k_+\,,\nonumber\\
   &&c_1|\gamma|^2\alpha^*k_- -i c_5 |\alpha|^2\gamma(k_3+k)\,,\nonumber\\
   &-&c_1|\gamma|^2\alpha^* (k_3-k) - ic_5 |\alpha|^2\gamma k_+
\Big\}\,,
\end{eqnarray}
\begin{eqnarray}
\lambdab_2(\vec{x},t)={2\e^{-i\vec{k}\cdot\vec{x}}\over a_0^{3/2} \e^{Ht}}{W_{+1/2,\nu}(z) \over [\alpha \gamma k_+ + \alpha^*\gamma^* k_-]}
   &\Big\{&
   c_7|\alpha|^2\gamma^*k_- + i c_3  |\gamma|^2\alpha(k_3-k)\,,\nonumber\\
   &-&c_7|\alpha|^2\gamma^*(k_3+k) + ic_3 |\gamma|^2\alpha k_+\,,\nonumber\\
   &&c_3|\gamma|^2\alpha^*k_- -i c_7 |\alpha|^2\gamma(k_3+k)\,,\nonumber\\
   &-&c_3|\gamma|^2\alpha^* (k_3-k) - ic_7 |\alpha|^2\gamma k_+
\Big\}\,,
\end{eqnarray}
\begin{eqnarray}
\lambdab_3(\vec{x},t)={2\e^{-i\vec{k}\cdot\vec{x}}\over a_0^{3/2} \e^{Ht}}{M_{-1/2,\nu}(z) \over [\alpha \gamma k_+ + \alpha^*\gamma^* k_-]}
   &\Big\{&
   c_6|\alpha|^2\gamma^*k_- + i c_2  |\gamma|^2\alpha(k_3+k)\,,\nonumber\\
   &-&c_6|\alpha|^2\gamma^*(k_3-k) + ic_2 |\gamma|^2\alpha k_+\,,\nonumber\\
   &&c_2|\gamma|^2\alpha^*k_- -i c_6 |\alpha|^2\gamma(k_3-k)\,,\nonumber\\
   &-&c_2|\gamma|^2\alpha^* (k_3+k) - ic_6 |\alpha|^2\gamma k_+
\Big\}\,,
\end{eqnarray}
\begin{eqnarray}
\lambdab_4(\vec{x},t)={2\e^{-i\vec{k}\cdot\vec{x}}\over a_0^{3/2} \e^{Ht}}{W_{-1/2,\nu}(z) \over [\alpha \gamma k_+ + \alpha^*\gamma^* k_-]}
   &\Big\{&
   c_8|\alpha|^2\gamma^*k_- + i c_4  |\gamma|^2\alpha(k_3+k)\,,\nonumber\\
   &-&c_8|\alpha|^2\gamma^*(k_3-k) + ic_4 |\gamma|^2\alpha k_+\,,\nonumber\\
   &&c_4|\gamma|^2\alpha^*k_- -i c_8 |\alpha|^2\gamma(k_3-k)\,,\nonumber\\
   &-&c_4|\gamma|^2\alpha^* (k_3+k) - ic_8 |\alpha|^2\gamma k_+
\Big\}\,.
\end{eqnarray}


\subsection{Case $a(t)=a_0 t$}

For the case $a(t)=a_0 t$, which represents the limit between the decelerated to the accelerated universe, equations (\ref{eqmotion21}) - (\ref{eqmotion24}) have the following linearly independent solutions in terms of the Bessel $J_{\nu}(z)$ and $Y_{\nu}(z)$ functions \cite{math}:
\beqq
\phi_1(t) = {2\sqrt{t}\over \alpha \gamma k_+ + \alpha^*\gamma^* k_-}&\bigg[& \Big(c_5\alpha^*\gamma^* k_- + i c_1|\gamma|^2( k_3 +k)\Big) J_{\nu_-}(z)\nonumber\\&+& \Big(c_7\alpha^*\gamma^*k_- + i c_3 |\gamma|^2(k_3+k)\Big)Y_{\nu_-}(z)\nonumber\\ &+&\Big(c_6\alpha^*\gamma^*k_- + ic_2 |\gamma|^2(k_3-k)\Big)J_{\nu_+}(z)\nonumber\\&+& \Big(c_8\alpha^*\gamma^*k_- + ic_4 |\gamma|^2(k_3-k)\Big)Y_{\nu_+}(z)\,\,\bigg]\,,\label{eq319}
\eeqq
\beqq
\phi_3(t) = {2\sqrt{t}\over \alpha \gamma k_+ + \alpha^*\gamma^* k_-}&\bigg[& \Big( c_1\alpha^*\gamma^* k_--i c_5|\alpha|^2(k_3 -k)\Big) J_{\nu_-}(z)\nonumber\\&+& \Big(c_3\alpha^*\gamma^*k_- - i c_7 |\alpha|^2(k_3-k)\Big)Y_{\nu_-}(z)\nonumber\\ &+&\Big(c_2\alpha^*\gamma^*k_- - ic_6 |\gamma|^2(k_3+k)\Big)J_{\nu_+}(z)\nonumber\\&+& \Big(c_4\alpha^*\gamma^*k_- - ic_8 |\gamma|^2(k_3+k)\Big)Y_{\nu_+}(z)\,\,\bigg]\,,\label{eq320}
\eeqq
where $\nu_\pm =(1/2)\sqrt{7-4k^2/a_0^2 \pm 4ik/a_0}$ and $z=mt$. 

The four independent solutions are:
\begin{eqnarray}
\lambda_1(\vec{x},t)={2\e^{i\vec{k}\cdot\vec{x}}\over a_0^{3/2} t}{J_{\nu_-}(z) \over [\alpha \gamma k_+ + \alpha^*\gamma^* k_-]}
   \left(\begin{array}{c}
   c_5|\alpha|^2\gamma^*k_- + i c_1  |\gamma|^2\alpha(k_3+k)\\
   -c_5|\alpha|^2\gamma^*(k_3-k) + ic_1 |\gamma|^2\alpha k_+\\
   c_1|\gamma|^2\alpha^*k_- -i c_5 |\alpha|^2\gamma(k_3-k)\\
   -c_1|\gamma|^2\alpha^* (k_3+k) - ic_5 |\alpha|^2\gamma k_+
\end{array}\right)\,, \label{sol5}
\end{eqnarray}
\begin{eqnarray}
\lambda_2(\vec{x},t)={2\e^{i\vec{k}\cdot\vec{x}}\over a_0^{3/2} t}{Y_{\nu_-}(z) \over [\alpha \gamma k_+ + \alpha^*\gamma^* k_-]}
   \left(\begin{array}{c}
   c_7|\alpha|^2\gamma^*k_- + i c_3  |\gamma|^2\alpha(k_3+k)\\
   -c_7|\alpha|^2\gamma^*(k_3-k) + ic_3 |\gamma|^2\alpha k_+\\
   c_3|\gamma|^2\alpha^*k_- -i c_7 |\alpha|^2\gamma(k_3-k)\\
   -c_3|\gamma|^2\alpha^* (k_3+k) - ic_7 |\alpha|^2\gamma k_+
\end{array}\right)\,, \label{sol6}
\end{eqnarray}
\begin{eqnarray}
\lambda_3(\vec{x},t)={2\e^{i\vec{k}\cdot\vec{x}}\over a_0^{3/2} t}{J_{\nu_+}(z) \over [\alpha \gamma k_+ + \alpha^*\gamma^* k_-]}
   \left(\begin{array}{c}
   c_6|\alpha|^2\gamma^*k_- + i c_2  |\gamma|^2\alpha(k_3-k)\\
   -c_6|\alpha|^2\gamma^*(k_3+k) + ic_2 |\gamma|^2\alpha k_+\\
   c_2|\gamma|^2\alpha^*k_- -i c_6 |\alpha|^2\gamma(k_3+k)\\
   -c_2|\gamma|^2\alpha^* (k_3-k) - ic_6 |\alpha|^2\gamma k_+
\end{array}\right)\,, \label{sol7}
\end{eqnarray}
\begin{eqnarray}
\lambda_4(\vec{x},t)={2\e^{i\vec{k}\cdot\vec{x}}\over a_0^{3/2} t}{Y_{\nu_+}(z) \over [\alpha \gamma k_+ + \alpha^*\gamma^* k_-]}
   \left(\begin{array}{c}
   c_8|\alpha|^2\gamma^*k_- + i c_4  |\gamma|^2\alpha(k_3-k)\\
   -c_8|\alpha|^2\gamma^*(k_3+k) + ic_4 |\gamma|^2\alpha k_+\\
   c_4|\gamma|^2\alpha^*k_- -i c_8 |\alpha|^2\gamma(k_3+k)\\
   -c_4|\gamma|^2\alpha^* (k_3-k) - ic_8 |\alpha|^2\gamma k_+
\end{array}\right)\,. \label{sol8}
\end{eqnarray}

For the anti-self-conjugate spinor $\lambdab$ we have:
\begin{eqnarray}
\lambdab_1(\vec{x},t)={2\e^{-i\vec{k}\cdot\vec{x}}\over a_0^{3/2} t}{J_{\nu_-}(z) \over [\alpha \gamma k_+ + \alpha^*\gamma^* k_-]}
   &\Big\{&
   c_5|\alpha|^2\gamma^*k_- + i c_1  |\gamma|^2\alpha(k_3-k)\,,\nonumber\\
   &-&c_5|\alpha|^2\gamma^*(k_3+k) + ic_1 |\gamma|^2\alpha k_+\,,\nonumber\\
   &&c_1|\gamma|^2\alpha^*k_- -i c_5 |\alpha|^2\gamma(k_3+k)\,,\nonumber\\
   &-&c_1|\gamma|^2\alpha^* (k_3-k) - ic_5 |\alpha|^2\gamma k_+
\Big\}\,,
\end{eqnarray}
\begin{eqnarray}
\lambdab_2(\vec{x},t)={2\e^{-i\vec{k}\cdot\vec{x}}\over a_0^{3/2} t}{Y_{\nu_-}(z) \over [\alpha \gamma k_+ + \alpha^*\gamma^* k_-]}
   &\Big\{&
   c_7|\alpha|^2\gamma^*k_- + i c_3  |\gamma|^2\alpha(k_3-k)\,,\nonumber\\
   &-&c_7|\alpha|^2\gamma^*(k_3+k) + ic_3 |\gamma|^2\alpha k_+\,,\nonumber\\
   &&c_3|\gamma|^2\alpha^*k_- -i c_7 |\alpha|^2\gamma(k_3+k)\,,\nonumber\\
   &-&c_3|\gamma|^2\alpha^* (k_3-k) - ic_7 |\alpha|^2\gamma k_+
\Big\}\,,
\end{eqnarray}
\begin{eqnarray}
\lambdab_3(\vec{x},t)={2\e^{-i\vec{k}\cdot\vec{x}}\over a_0^{3/2} t}{J_{\nu_+}(z) \over [\alpha \gamma k_+ + \alpha^*\gamma^* k_-]}
&\Big\{&
   c_6|\alpha|^2\gamma^*k_- + i c_2  |\gamma|^2\alpha(k_3+k)\,,\nonumber\\
   &-&c_6|\alpha|^2\gamma^*(k_3-k) + ic_2 |\gamma|^2\alpha k_+\,,\nonumber\\
   &&c_2|\gamma|^2\alpha^*k_- -i c_6 |\alpha|^2\gamma(k_3-k)\,,\nonumber\\
   &-&c_2|\gamma|^2\alpha^* (k_3+k) - ic_6 |\alpha|^2\gamma k_+
\Big\}\,,
\end{eqnarray}
\begin{eqnarray}
\lambdab_4(\vec{x},t)={2\e^{-i\vec{k}\cdot\vec{x}}\over a_0^{3/2} t}{Y_{\nu_+}(z) \over [\alpha \gamma k_+ + \alpha^*\gamma^* k_-]}
  &\Big\{&
   c_8|\alpha|^2\gamma^*k_- + i c_4  |\gamma|^2\alpha(k_3+k)\,,\nonumber\\
   &-&c_8|\alpha|^2\gamma^*(k_3-k) + ic_4 |\gamma|^2\alpha k_+\,,\nonumber\\
   &&c_4|\gamma|^2\alpha^*k_- -i c_8 |\alpha|^2\gamma(k_3-k)\,,\nonumber\\
   &-&c_4|\gamma|^2\alpha^* (k_3+k) - ic_8 |\alpha|^2\gamma k_+
\Big\}\,.
\end{eqnarray}

\subsection{Case $a(t)=a_0 \sqrt{t}$}

For the case $a(t)=a_0 \sqrt{t}$, which represents a radiation dominated universe, the equations (\ref{eqmotion21}) - (\ref{eqmotion24}) has the following linearly independent solutions\footnote{There are other independent solutions written in terms of integral equations, but we are omitting these solutions here (see ref. \cite{heun,ronveaux} for more details).} in terms of the Heun B functions, denoted here by $B_{a,b,\mu,\nu}(z)$, that are solutions of the Heun biconfluent equation \cite{heun,ronveaux}:
\beqq
\phi_1(t) = {2 t \e^{imt}\over \alpha \gamma k_+ + \alpha^*\gamma^* k_-}&\bigg[& \Big(c_3\alpha^*\gamma^* k_- +ic_1|\gamma|^2( k_3 +k)\Big) B_{2,0,\mu,\nu_-}(z) \nonumber\\
&+&\Big(c_4\alpha^*\gamma^*k_-+ ic_2|\gamma|^2( k_3 -k)\Big) B_{2,0,\mu,\nu_+}(z) \bigg]\,,
\eeqq
\beqq
\phi_3(t) = {2 t \e^{imt}\over \alpha \gamma k_+ + \alpha^*\gamma^* k_-}&\bigg[& \Big(c_1\alpha^*\gamma^* k_- -ic_3|\alpha|^2( k_3 -k)\Big) B_{2,0,\mu,\nu_-}(z) \nonumber\\
&+&\Big(c_2\alpha^*\gamma^*k_-- ic_4|\alpha|^2( k_3 +k)\Big) B_{2,0,\mu,\nu_+}(z) \bigg]\,,
\eeqq
where $\mu=2ik^2/a_0^2 m$, $\nu_\pm = \pm(2-2i)k/a_0\sqrt{m}$ and $z=(-1+i)\sqrt{mt}$. 

The two independent solutions are:
 \begin{eqnarray}
\lambda_1(\vec{x},t)={2\e^{i\vec{k}\cdot\vec{x}}t^{1/4} \over a_0^{3/2}}{B_{2,0,\mu,\nu_-}(z)\over [\alpha \gamma k_+ + \alpha^*\gamma^* k_-]}
   \left(\begin{array}{c}
  \e^{imt}[ c_3|\alpha|^2\gamma^*k_- + i c_1  |\gamma|^2\alpha(k_3+k)]\\
  \e^{-imt}[ -c_3|\alpha|^2\gamma^*(k_3-k) + ic_1 |\gamma|^2\alpha k_+]\\
   \e^{imt}[c_1|\gamma|^2\alpha^*k_- -i c_3 |\alpha|^2\gamma(k_3-k)]\\
   \e^{-imt}[-c_1|\gamma|^2\alpha^* (k_3+k) - ic_3 |\alpha|^2\gamma k_+]
\end{array}\right)\,, \label{sol9}
\end{eqnarray}
 \begin{eqnarray}
\lambda_2(\vec{x},t)={2\e^{i\vec{k}\cdot\vec{x}}t^{1/4} \over a_0^{3/2}}{B_{2,0,\mu,\nu_+}(z)\over [\alpha \gamma k_+ + \alpha^*\gamma^* k_-]}
   \left(\begin{array}{c}
  \e^{imt}[ c_4|\alpha|^2\gamma^*k_- + i c_2  |\gamma|^2\alpha(k_3-k)]\\
  \e^{-imt}[ -c_4|\alpha|^2\gamma^*(k_3+k) + ic_2 |\gamma|^2\alpha k_+]\\
   \e^{imt}[c_2|\gamma|^2\alpha^*k_- -i c_4 |\alpha|^2\gamma(k_3+k)]\\
   \e^{-imt}[-c_2|\gamma|^2\alpha^* (k_3-k) - ic_4 |\alpha|^2\gamma k_+]
\end{array}\right)\,. \label{sol10}
\end{eqnarray}

For the anti-self-conjugate spinor $\lambdab$ we have:
 \begin{eqnarray}
\lambdab_1(\vec{x},t)={2\e^{-i\vec{k}\cdot\vec{x}}t^{1/4} \over a_0^{3/2}}{B_{2,0,\mu,\nu_-}(z)\over [\alpha \gamma k_+ + \alpha^*\gamma^* k_-]}
   &\Big\{&
  \e^{imt}[ c_3|\alpha|^2\gamma^*k_- + i c_1  |\gamma|^2\alpha(k_3-k)]\,,\nonumber\\
  &&\e^{-imt}[ -c_3|\alpha|^2\gamma^*(k_3+k) + ic_1 |\gamma|^2\alpha k_+]\,,\nonumber\\
   &&\e^{imt}[c_1|\gamma|^2\alpha^*k_- -i c_3 |\alpha|^2\gamma(k_3+k)\,,\nonumber]\\
   &&\e^{-imt}[-c_1|\gamma|^2\alpha^* (k_3-k) - ic_3 |\alpha|^2\gamma k_+]
\Big\}\,,\nonumber
\end{eqnarray}
 \begin{eqnarray}
\lambdab_2(\vec{x},t)={2\e^{-i\vec{k}\cdot\vec{x}}t^{1/4} \over a_0^{3/2}}{B_{2,0,\mu,\nu_+}(z)\over [\alpha \gamma k_+ + \alpha^*\gamma^* k_-]}
 &\Big\{&
  \e^{imt}[ c_4|\alpha|^2\gamma^*k_- + i c_2  |\gamma|^2\alpha(k_3+k)]\,,\nonumber\\
  &&\e^{-imt}[ -c_4|\alpha|^2\gamma^*(k_3-k) + ic_2 |\gamma|^2\alpha k_+]\,,\nonumber\\
   &&\e^{imt}[c_2|\gamma|^2\alpha^*k_- -i c_4 |\alpha|^2\gamma(k_3-k)]\,,\nonumber\\
   &&\e^{-imt}[-c_2|\gamma|^2\alpha^* (k_3+k) - ic_4 |\alpha|^2\gamma k_+]
\Big\}\,.\nonumber
\end{eqnarray}

\section{Elko spinor in cosmology}

Some peculiar features of the Elko field have been used in order to extract physical information about cosmological scenarios. For instance, a quite interesting mass upper bound may be found in trying to use Elko fields as dark matter driving inflation \cite{BOE1}. In this section we shall consider a simple model of dark energy, instead, driven by the Elko spinor. 

As has been done in recent works \cite{BOE3,BOE4,BOE6,BOE7,WEI}, usually the spinor field is factored out to a real homogeneous scalar field, $\lambda\equiv \varphi(t)\xi$, with the temporal dependence only in $\varphi(t)$, the same for all components and $\xi$ representing a constant and normalized Elko spinor. Due to the homogeneity of the field ($\partial_i \lambda =0$), the equation (\ref{eqmotion}) for $\varphi(t)$ is substantially simplified,
\be
\ddot{\varphi}+3H\dot{\varphi}-{3\over 4}H^2\varphi+m^2\varphi=0\,,\label{eqphi}
\ee
where $H=\dot{a}/a$. We shall comment on this simplification at the end of this Section. The pressure and energy density are given by \cite{BOE6}
\be
p_\varphi={1\over 2}\dot{\varphi}^2-{1\over 2}m^2\varphi^2-{3\over 8}H^2\varphi^2-{1\over 4}\dot{H}\varphi^2-{1\over 2}H\varphi\dot{\varphi}\,,\label{press}
\ee
\be
\rho_\varphi={1\over 2}\dot{\varphi}^2+{1\over 2}m^2\varphi^2+{3\over 8}H^2\varphi^2\,.\label{rho}
\ee
The Friedmann equations for $H(t)$ can be written as
\be
H^2={8\pi G\over 3}\rho\,\hspace{1cm}\dot{H}=-{4\pi G}(\rho + p)\,, \label{eqfried}
\ee
from which follows the conservation equation
\be
\dot{\rho}+3H(\rho + p)=0\,,\label{conservation}
\ee
where $\rho$ and $p$ stands for the total energy density and total pressure of all the matter fields present in the model.

In this simplified model we will consider that all the material content of the universe is the Elko spinor satisfying a dark energy equation of state $p=-\rho$, thus the Friedmann equations reduces to
\be
H={\dot{a}\over a}=\pm\Bigg({8\pi G\over 3}\rho_\varphi\Bigg)^{1/2}\,,\hspace{1cm} \dot{H}=0\,,
\ee
furthermore we have $\dot{\rho}_\varphi =0$, so that $\rho_\varphi$ is a constant, implying a de Sitter evolution $a(t)=a_0\e^{Ht}$, thus we can use the solutions obtained in Section 3.1. 

Before we proceed, let us examine the restriction imposed by the dark energy equation of state $p_\varphi=-\rho_\varphi$. By using (\ref{press}) and (\ref{rho}) we find
\be
\dot{\varphi}^2-{1\over 2}H\varphi\dot{\varphi}=0\,,
\ee
whose solutions are of two types, namely static or dynamic,
\be
\varphi(t)=\bar{\varphi} \hspace{0.4cm}\textrm{(static)}\,,\hspace{1cm} \varphi(t)=\varphi_0\e^{Ht/2}\hspace{0.4cm}\textrm{(dynamic)}\,.\label{solsSD}
\ee

As it can be read from Eqs. (\ref{sol1})-(\ref{sol4}), the obtained solutions are already in the time factored form\footnote{Indeed, for all the three cases analysed here, the solutions given by Eqs. (\ref{sol1}-\ref{sol4}), (\ref{sol5}-\ref{sol8}) and (\ref{sol9}-\ref{sol10}) are already in the time factored form, which justifies the use of the decomposition $\lambda=\varphi(t)\xi$ in recent works \cite{BOE3,BOE4,BOE6,BOE7,WEI}.}. Therefore, to be able to use the above equations we shall only to apply some limit in order to get the homogeneous solution. This limit can be achieved by taking into account that the inhomogeneity comes from spatial derivatives, giving rise to momentum dependent terms. Hence, the homogeneous limit is obtained by taken $k_{1,2,3}$ to zero carefully. For the argument, it is always possible to restrict the momentum to one direction and then take this momentum to vanish. Once this limit is performed, two things happen: 1) the spinorial part of the solution becomes homogeneous and the normalization can be imputed to the integration constants; 2) the Whittaker $k\rightarrow 0$ (or, correspondingly $z\rightarrow 0$) limits are in order. As the Whittaker limit $M_{\sigma,\nu}(z\rightarrow 0)$ depends on the $\sigma$ and $\nu$ index, we must explore all the possibilities.  

Let us start investigating the solution (\ref{sol1}) in the homogeneous limit. It is easy to see that in this case we have 
\be
\lambda_1(t)=\frac{2}{a_0^{3/2}}M_{+1/2,\nu}(z\rightarrow 0)e^{-Ht}\xi,\label{ufa1}
\ee where $\xi$ stands for the constant homogeneous spinorial part of $\lambda_1$ (which is completely irrelevant to this application). The unique $z\rightarrow 0$ limit allowed for this specific Whittaker function occurs when $2\nu \neq -1,-2,\ldots$. Supposing this is the case we have $M_{+1/2,\nu}\rightarrow z^{\nu+1/2}$. Hence, writing the solution as $\lambda_1(t)=\varphi(t)\xi$, bearing in mind that $z=2ik/a_0He^{Ht}$ and absorbing the constant part in $\varphi_0$ one gets 
\be 
\varphi=\varphi_0 \e^{-(\nu+3/2)t}. \label{ufa2}
\ee Comparing the solution (\ref{ufa2}) with the static solution (\ref{solsSD}) we see that it entails $2\nu=-3$ contradicting the hypothesis for this limit validity. Thus this solution cannot describe the static situation. To fulfil the dynamic solution of (\ref{solsSD}) we must have $\nu=-2$, but as $\nu>0$ by definition ($\nu=\sqrt{3-m^2/H^2}$), we shall disregard this solution.

The solution given by (\ref{sol3}), $\lambda_3$, is more interesting. In fact, the time dependent part in this case can be recast as 
\be
\varphi(t)=\varphi_0 e^{(\nu-3/2)Ht}.\label{ufa3}
\ee By comparing Eq. (\ref{ufa3}) with (\ref{solsSD}) we see that the dynamical solution requires $\nu=2$, which is mathematically acceptable, but it leads to $m^2<0$,  a physically unacceptable condition. It is interesting, however, that massive ghosts solutions to an Einstein-Cartan-Dirac system present the very same behavior \cite{FANT}. The static case, on the other hand, can be reached if $\nu=3/2$, leading to an Elko mass given by $m=\frac{\sqrt{3}}{2}H$, providing a physically acceptable solution. In fact, a static solution for the field $\varphi$ can be directly obtained from (\ref{eqphi}) if the above condition on the mass is set. By analyzing the energy density (\ref{rho}) with a static field $\varphi$ we see that in this case it reproduces exactly a cosmological constant term in the Friedmann equation (\ref{eqfried}). 

It can be straightforwardly verified that the remain cases ($\lambda_2$ and $\lambda_4$) do not contain any novelty, leading to nonphysical solutions or reproducing the static case for $m=\frac{\sqrt{3}}{2}H$.

We would like to conclude this section by tracing some comments on the homogeneity simplification in the spinor solutions. In fact, when this is the case, we have arrived at a relationship between the spinor mass and $H$ so that the solution can be applied to this cosmological scenario (also simplified). The net effect of considering the non-homogeneous case is, probably, the obtention of a more general vinculum, this time regarding the mass, the Hubble parameter and the momentum. This dispersion relation-like constraint may, eventually, lead to new possible cosmological applications, but there is no guarantee that it is in fact physically appealing. 

\section{Concluding remarks}

In the present work we have studied the evolution of Elko spinors in a flat Friedmann-Robertson-Walker background finding exact solutions for three different models of expansion, namely a de Sitter, linear and radiation. A very interesting aspect of the solutions we have found is that, contrary to the solutions of the Dirac equation in a spatially flat Friedmann-Robertson-Walker spacetimes \cite{barut} where the first two components are coupled to the last two, the equations for the first two components of the Elko spinors are independent of the third and fourth, as can be seen in Eq. (\ref{eqmotion1}). 

Still comparing with the Dirac case, we see that the solution for the temporal part of the spinor is totally different in the three cases examined here. For the de Sitter evolution, the Dirac case gives solutions in terms of the Bessel functions, while here we obtain solutions in terms of the Whittaker functions. For both linear and radiation expansion the solutions in the Dirac case are given by means of Whittaker functions, while here we obtain the Bessel functions for the linear evolution and the complicated Heun biconfluent functions as solutions for the radiation. These behaviors illustrate some of the differences between the Dirac and the Elko spinors. 

We have investigated a cosmological setting where an homogeneous Elko spinor acts as dark energy in a de Sitter background. It is shown that there are two solutions for this case. A dynamic one, where a constraint in the mass parameter indicates a non-physical scenario and a static solution, in which the spinor field works as an effective cosmological constant. It is important to emphasize that the results of such an application are mainly due to the maintenance of the state equation $p=-\rho$ leading to a de Sitter expansion. Other types of equation of state, giving different expansion rates are not excluded in principle, but dealing with the resulting coupled differential equations system is certainly a difficult endeavour.  

We shall finish remembering that here our analysis was restricted to the flat Friedmann-Robertson-Walker geometry. Generalizations involving the parabolic and hyperbolic curved backgrounds may be achieved. Some applications involving particle creation in more general spacetimes are under investigation.


\acknowledgments

SHP is grateful to CNPq for the financial support, process number 477872/2010-7. JMHS thanks CNPq (482043/2011-3, 308623/2012-6) for partial support. The authors are grateful to Professor Rold\~ao da Rocha for careful reading of the manuscript. 


\end{document}